\documentclass[12pt]{article}
\usepackage[utf8]{inputenc}
\usepackage{amsmath,latexsym,amssymb,graphicx}
\usepackage{hyperref}
\usepackage{caption}
\usepackage{subcaption}
\usepackage{cite}
\hypersetup{
    colorlinks=true,
    filecolor=magenta,      
    urlcolor=cyan,
    citecolor=blue
}
\usepackage{authblk}

\DeclareMathOperator{\arcot}{arccot}
\DeclareMathOperator{\Li2}{Li_2}
\DeclareMathOperator{\Lin}{Li_n}

\begin{document}

\title{Massless fermions localization on domain walls}
\author[$\dagger$]{Rommel Guerrero}
\author[$\dagger$]{R. Omar Rodriguez}
\author[$\star$]{Francisco Carreras}
\affil[$\dagger$]{Grupo de Investigaciones en F\'isica, Escuela de F\'isica y Matem\'atica, Facultad de Ciencias, Escuela Superior Polit\'ecnica de Chimborazo, EC060155-Riobamba, Ecuador.}
\affil[$\star$]{ Escuela de F\'isica y Matem\'atica, Facultad de Ciencias, Escuela Superior Polit\'ecnica de Chimborazo, EC060155-Riobamba, Ecuador.}

\maketitle

\begin{abstract}
Massless fermions on scalar domain walls are considered. Two walls are established, corresponding to 5-dimensional static spacetime asymptotically Anti de-Sitter, differentiated  by the symmetry around the wall, and in each case massless chiral fermions are coupled to the wall by a Yukawa term. We identify a normalizable state associated to the migration of fermions toward the edge of the wall. This effect is generated by the competition between the Yukawa interaction and the gravitational repulsion on the matter fields, and it is independent of the $Z_2$ symmetry of the wall.\\


\end{abstract}
\maketitle

\section{Introduction}\label{intro}

A brane world corresponds to a 4-dimensional hypersurface embedded in a  high dimensional bulk, on which gravity is localized \cite{Randall:1999ee,Randall:1999vf}. In the most general case, the 4-dimensional  sector can be generated by a scalar domain wall interpolating between two asymptotic Anti de-Sitter (AdS) vacuums, with different cosmological constants  at each side of the wall, i.e., a solution to the coupled Einstein-Klein Gordon system with a potential with  a spontaneously broken
discrete symmetry \cite{Skenderis:1999mm, Gremm:1999pj,DeWolfe:1999cp, Melfo:2002wd, Guerrero:2002ki, Bazeia:2003aw,CastilloFelisola:2004eg,Guerrero:2005xx,Guerrero:2005aw}. The gravity fluctuations of this spacetime depend on the wall’s thickness, and are characterized by both a zero mode localized on the wall, and a spectrum of massive states propagating freely. In the thin wall limit \cite{Guerrero:2002ki,Pantoja:2003zr}, the gravitational potential can be calculated, and Newtonian gravity is recovered on the 4-dimensional sector of the scenario.

To localize the matter field, a Yukawa coupling between the bulk fermions and the scalar field of the wall is required; however, in this case, only one chiral massless state (left or right) is localized, while the other is repelled by the gravitational field of the wall \cite{Rubakov:1983bb, ArkaniHamed:1999dc, Bajc:1999mh, RandjbarDaemi:2000cr, Dubovsky:2000am, Kehagias:2000au,  Ringeval:2001cq,Koley:2004at,Melfo:2006hh, Guerrero:2006gj,Liu:2009dw,Liu:2009dwa,Liu:2009ve,Bazeia:2017zye}. Remarkably, we determined that the minimum bound of the Yukawa constant, obtained from  conditions of normalization, is not sufficient to confine the fermion in 4-dimensions with maximum probability on the scalar wall.

On the other hand, it is well known that in absence of $Z_2$ symmetry, the gravitational field of the brane shifts the fermion toward the 5-dimensional sector with greater curvature \cite{Guerrero:2006gj}. Here, we show that another shift-like effect exists, regardless of the reflection symmetry of the scenario. For some values of the Yukawa coupling constant, the chiral fermion can be shifted  simultaneously at each side of the wall, i.e., the bound state exhibits two relative maximums close to the wall and hence the probability of finding the particle is greater around the wall rather than on it. 

To rule out the relationship between the symmetry of the scenario and the fermion's migration, we consider two models differentiated by their reflection symmetry. The first one corresponds to the member  with $Z_2$ symmetry of a family of asymmetric domain walls \cite{Gremm:1999pj}, while the second one is an intrinsically asymmetric wall family \cite{CastilloFelisola:2004eg,Melfo:2006hh}. In both cases, when the fermion satisfies the normalization condition and it is distributed (centered) around the wall, two peaks at each side of the wall in the probability profile of the fermion are obtained. For the other cases, where only the normalization condition is satisfied, the fermion is always confined outside of the wall.

\section{Set-up}\label{S2}
Consider the coupled Einstein scalar field system in 5-dimensions (latin index $a,b=0,\cdots,4$; greek index $\mu,\nu=0,\cdots,3$)
 \begin{equation}\label{Action}
S=\int dx^5 \sqrt{g}\left[\frac{1}{2}R-\frac{1}{2}g^{ab}\nabla_a\phi\nabla_b\phi-V(\phi)\right], 
 \end{equation}
where the metric is given by 
 \begin{equation}
ds^2=e^{2A(y)}\eta_{\mu\nu}dx^\mu dx^\nu+dy^2, \label{metric}
 \end{equation}
which corresponds  to a spacetime with plane parallel symmetry whose  energy density is determined by
\begin{equation}
    \rho=-3(2A^{\prime 2}+A^{\prime\prime}).
\end{equation}
Here and in the other sections prime denotes the derivative with respect to $y$.
 
We are interested in AdS${}_5$ domain wall solutions, scenarios for which geometry is determined by a scalar field $\phi$ interpolating between the negative minima of the potential $V(\phi)$. It is well known that the full spectrum of the gravity fluctuations of these scenarios is characterized by a zero mode localized on the wall, $\psi_{\mathrm{g}}\sim e^{2 A(y)}$,
and a continuum of massive modes propagating freely through the whole 5-dimensional bulk; see \cite{CastilloFelisola:2004eg} for details. 

The chiral massless fermions can also be localized on the wall by a Yukawa coupling between the bulk fermions and the scalar field, $\lambda\bar\Psi\phi\Psi$ \cite{Melfo:2006hh}. To evidence this, the Dirac equation for the 5-dimensional spinor in the background (\ref{metric}) must be considered  
\begin{equation}
    \Gamma^a\nabla_a\Psi(x,y)=\lambda\phi(y)\Psi(x,y).
\end{equation}

 To obtain the coordinates representation of the motion equation, it is common to rewrite $\Psi(x,y)$ in terms of its chiral components left (L) and right (R), and factor its spatial degrees of freedom, as shown below  
\begin{equation}
    \Psi(x,y)=\Psi_{\text{L}}(x)\psi_{\text{L}}(y)+\Psi_{\text{R}}(x)\psi_{\text{R}}(y),
\end{equation}
where $\Psi_{\text{R}}^{\text{L}}(x)\equiv\pm\gamma^5\Psi_{\text{R}}^{\text{L}}(x)$, which satisfies the massless 4-dimensional Dirac equation
\begin{equation}\label{Dirac4}
    i\gamma^\mu\partial_\mu\Psi_{\text{R}}^{\text{L}}(x)=0 .
\end{equation}
Thus, for $\psi_{\text{R}}^{\text{L}}(y)$ we have 
\begin{equation}
    \left(\partial_y+2A^\prime(y)\pm\lambda \phi(y) \right)\psi_{\text{R}}^{\text{L}}(y)=0 ; \label{hatu}
\end{equation}
for which %solution is determined by
 \begin{equation}\label{hs}
    \psi_{\text{R}}^{\text{L}}(y)=e^{-2A(y)\mp\lambda\int{ \phi(y) }\ dy}.
\end{equation}

For a domain wall solution, $e^A$ is an integrable and asymptotically vanishing function, such that,  $e^{-A}\sim e^{k|y|}$ as $| ky|\gg 1$, with $k=\sqrt{|\Lambda|/6}$, i.e., gravitation produces a repulsive effect on fermions. Hence, for the appropriate values of $\lambda$ one of the  $\psi_{\text{R}}^{\text{L}}$ can be normalized.  In particular, for $\psi_{\text{L}}$ in the large $y$ limit, it is possible to see that
 \begin{equation}\label{h}
    \psi_{\text{L}}(y)\longrightarrow e^{(2k-\lambda\phi_0)|y|} ,
\end{equation}
where $\phi(y=\pm\infty)=\pm\phi_0$ has been considered. Therefore 
 \begin{equation}
\lambda > \frac{2k}{\phi_0}\ \equiv \lambda_1\label{cota1}
 \end{equation}
 in order to cancel the repulsive effect generated by gravity.  

The previous restriction for the Yukawa coupling is a necessary condition, but not sufficient to ensure the confinement of $\psi_{\text{L}}$ on the wall. For some values of $\lambda$, we found solutions for $\psi_{\text{L}}$ with two peaks at each sides of the wall, generated by the gravitational competition between the geometry of the spacetime and the Yukawa interaction.

Let $\lambda_1$ be the Yukawa critical constant from which (\ref{hs}) is normalizable. Moreover, let $\lambda_2$ be the Yukawa threshold from which the maximum of (\ref{hs})  is on the wall ($\lambda_2>\lambda_1$). Then, for $\lambda<\lambda_1$, the fermion is expelled from the thick brane by the gravitational field of the wall; while for $\lambda>\lambda_2$ the fermion is localized on the wall. On the other hand, the sector $\lambda_1<\lambda<\lambda_2$ corresponds to a bounded probability distribution with three critical points: a pair of maximums around the wall and a minimum on it. So, the chiral zero mode of the matter fields is repelled by the thick brane and keeps orbiting around it. 

The critical bound $\lambda_2$ can be determined by a saddle point evaluation of (\ref{hs}). By assuming that $\psi_{\text{L}}$ is peaked near the brane, $y=0$,  and that $\psi_{\text{L}}^{\prime}(0)=0$ and $\psi_{\text{L}}^{\prime\prime}(0)<0$, the following is obtained
 \begin{equation}
\int_{-\infty}^{\infty} dy\  \psi_{\text{L}}(y)\simeq\left[\frac{2\pi}{-2|A^{\prime\prime}(0)|+ \lambda\phi^\prime(0)}\right]^{1/2}\ \psi_{\text{L}}(0) .
 \end{equation}
Therefore, to preserve it as a real amount, the following is required 
 \begin{equation}
\lambda>2\frac{|A^{\prime\prime}(0)|}{\phi^\prime(0)}\equiv\lambda_2 \label{cota2}
 \end{equation}

Now, to determinate the hierarchy between $\lambda_1$ and $\lambda_2$ we will consider two specific solutions to the Einstein-Klein Gordon system (\ref{Action}), differentiated by the reflection symmetry of each scenario.

\section{Symmetric walls}\label{S3}
A solution to the coupled system (\ref{Action}) is given by
 \begin{equation}
A(y)=-\delta\ln\cosh\frac{\alpha y}{\delta} ,
 \end{equation}
 \begin{equation}
\phi(y)=\phi_0\arctan\sinh\frac{\alpha y}{\delta} ,\qquad \phi_0=\sqrt{3\delta} ,
 \end{equation}
and
 \begin{equation}
V(\phi)= 3\alpha^2\left[\left(\frac{1}{2\delta}+2\right)\cos^2\frac{\phi}{\phi_0}-2\right].
 \end{equation}

This solution was obtained in \cite{Gremm:1999pj} and corresponds to a two-parameter domain wall interpolating asymptotically between two AdS${}_5$ spacetime, with a cosmological constant given by $\Lambda=-6\alpha^2$. Also, it is well known that the zero mode of gravity fluctuations is normalizable, and hence standard gravitation can be recovered in 4-dimensions. 

With regard to massless fermions, from (\ref{hs})
\begin{eqnarray}
    &&\psi_{\text{L}}\sim \cosh^{2\delta}\left(\alpha y/\delta\right)\exp\left[\lambda\frac{\delta\phi_0}{\alpha}\left(\frac{\alpha y}{\delta}(2\arcot e^{\alpha y/\delta}+\arctan\sinh\alpha y/\delta)\right)\right]\nonumber\\&&\quad\qquad\qquad\qquad \times \exp\left[i\lambda\frac{\delta\phi_0}{\alpha}\left(\Li2(-i e^{-\alpha y/\delta})-\Li2(i e^{-\alpha y/\delta})\right)\right]
\end{eqnarray}
where $\Lin$ is the polylogarithm function of order $n$ given by
\begin{equation}
    \Lin(\tau)\equiv\sum_{k=1}^\infty\frac{\tau^k}{k^n} ;
\end{equation}
and from (\ref{cota1}) and (\ref{cota2}) we find
 \begin{equation}
\lambda_1=\frac{4\alpha}{\pi\sqrt{3\delta}} 
 \end{equation}
 and 
  \begin{equation}
\lambda_2=\frac{\pi}{2}\lambda_1 ,
 \end{equation}
 respectively.
 
\begin{figure}[h!]
        \centering
        \begin{subfigure}{0.45\textwidth} 
            \includegraphics[width=\textwidth]{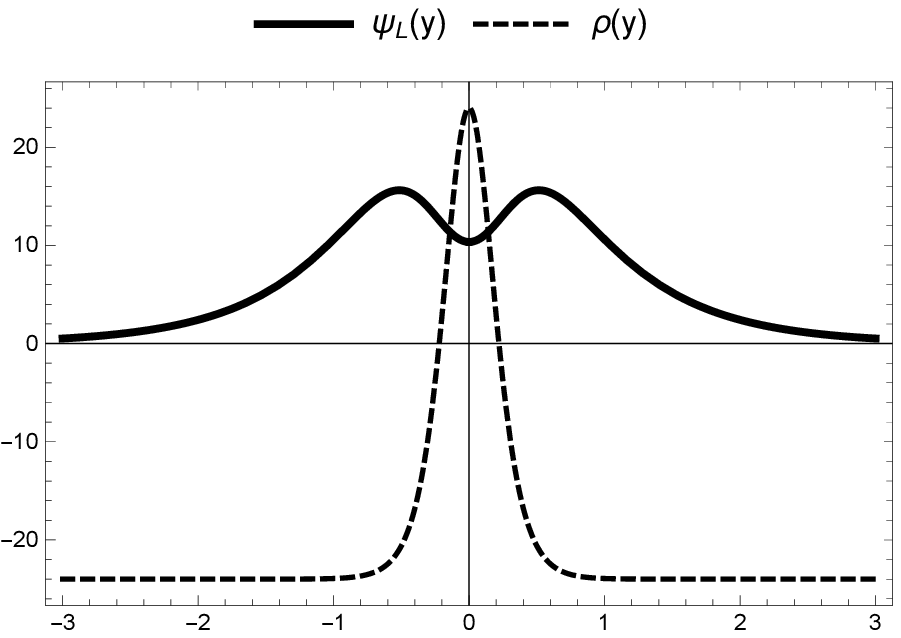}
        \end{subfigure}       
        \begin{subfigure}{0.45\textwidth} 
     \includegraphics[width=\textwidth]{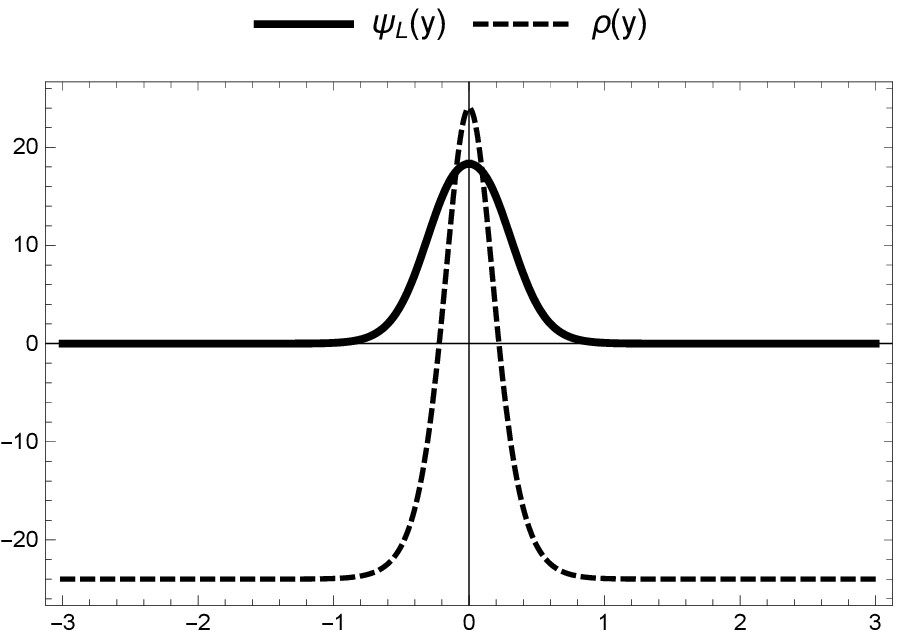}
        \end{subfigure}
        \caption{Massless fermion $\psi_{\text{L}}$  and energy density $\rho$  for $\lambda_1<\lambda<\lambda_2$ (left) and $\lambda>\lambda_2$ (right). }\label{plotrhos}
    \end{figure}

Fig.\ref{plotrhos} shows the fermions profile for some values of $\lambda$. We observed that for $\lambda>\lambda_2$, $\psi_{\text{L}}$ is normalizable, with maximum probability on the wall. On the other hand, if $\lambda_1<\lambda<\lambda_2$, 
a change occurs in the behavior of $\psi_{\text{L}}$; in this case, the competition between gravitational interaction and Yukawa coupling divides the maximum of profile into two maximums around the wall; i.e., the fermions prefer to orbit around the structure.

\section{Asymmetric walls}\label{S4}

Before considering a thick domain wall with different cosmological constants at each side of the wall, $\Lambda_+$ and $\Lambda_-$, it is convenient to estimate the asymptotic behavior of $\psi_{\text{L}}$ in this scenario, i.e.,
  \begin{equation}\label{PsiLA} 
    \psi_{\text{L}}(y)\longrightarrow \Theta(y) e^{(2k_{{}_+}-\lambda_{{}_+}|\phi_{{}_+}|)y} + \Theta(-y) e^{-(2k_{{}_-}-\lambda_{{}_-}|\phi_{{}_-}|)y} ,
\end{equation}
where $\phi(y=\pm\infty)=\phi_\pm$ and $k_\pm=\sqrt{|\Lambda_\pm|/6}$ have been considered. 

In order to obtain a normalizable solution for $\psi_{\text{L}}$, from (\ref{PsiLA}) reads
 \begin{equation}\label{lambdaY}
\lambda >\text{max}\{\lambda_{1+}, \lambda_{1-} \}\quad ,\quad \lambda_{1\pm}=\frac{2k_\pm}{|\phi_\pm|} .
 \end{equation}
Thus, two critical Yukawa constants associated to $\Lambda_\pm$ are obtained, which are the result of the lack of $Z_2$ symmetry. 

 Now, suppose that $\lambda_{1-}>\lambda_{1+}$ and that $\lambda_{1-}>\lambda_2>\lambda_{1+}$. In this case, although $\lambda_2$ is greater than one of the asymptotic constants, the migratory effect may not be present because $\lambda_2 < \lambda_{1-}$. On the other hand, there is the possibility  that $\lambda_2>\lambda_{1-}$  and even though the fermion is normalizable, it is more likely to orbit around the wall.  To analyze this, let us consider one specific solutions for the domain wall.
 
An asymmetric solution is determined by
\begin{equation}
A(y)=-\delta \exp \left(-2\ e^{-\alpha y/\delta}\right)+\delta \text{Ei} \left(-2\ e^{-\alpha y/\delta}\right)+\beta y \label{metrica},
\end{equation}
where $\alpha$ and $\delta$ are positive constants and $\text{Ei}$ is the exponential integral  given by
\begin{equation}
\text{Ei}(u)=-\int_{-u}^\infty e^{-t}/t \ dt.
\end{equation}
The scalar field and potential for this scenario are
\begin{equation}
\phi(y)=\phi_0\exp \left(-\ e^{-\alpha y/\delta}\right)-\epsilon
\label{fielda}
\end{equation}
and
\begin{eqnarray}
&&V(\phi)=18\left[\alpha\frac{(\phi+\epsilon)}{\phi_0^2}\ln\frac{(\phi+\epsilon)^2}{\phi_0^2}\right]^2-6\left[\alpha\frac{(\phi+\epsilon)^2}{\phi_0^2}\left(1-\ln\frac{(\phi+\epsilon)^2}{\phi_0^2}\right)\right]^2\nonumber\\&&\qquad\quad+6\beta\left[2\alpha\frac{(\phi+\epsilon)^2}{\phi_0^2}\left(1-\ln\frac{(\phi+\epsilon)^2}{\phi_0^2}\right)-\beta\right].\label{potentiala}
\end{eqnarray}

This solution was reported in \cite{CastilloFelisola:2004eg} and it corresponds to  a two-parameter family of plane symmetric static domain walls without $Z_2$ symmetry. The scalar field  interpolates between two asymptotic vacua, $\phi_-=-\epsilon$ for $y<0$ and $\phi_+=\phi_0-\epsilon$ for $y>0$, with different cosmological constants
\begin{equation}
{\Lambda}_-=-6\beta^2,\qquad {\Lambda}_+=-6(\alpha-\beta)^2,\qquad 0<\beta/\alpha<1\label{ccosmological2} .
\end{equation}

Now, equation (\ref{hs}) provides that
\begin{eqnarray}
&& \psi_{\text{L}}\sim\exp\lambda\left[\frac{\delta}{\alpha}\phi_0\text{Ei}\left(-\ e^{-\alpha y/\delta}\right)+\epsilon y\right]
\nonumber\\
&& \qquad\times\exp\left[2\delta \exp(-2e^{-\alpha y/\delta})-2\delta\text{Ei}\left(-2\ e^{-\alpha y/\delta}\right)-2\beta y\right] .
\end{eqnarray}
Notice that, constants $\epsilon$ and $\beta$ fix the asymptotic values of the scalar field and the cosmological constants, in such a way that, neither for the fermion nor for the graviton it is possible to find normalizable solutions as $\epsilon=\beta=0$, because $\phi_-=\Lambda_-=0$.

\begin{figure}[b!]
\centering
\includegraphics[width=0.5\textwidth]{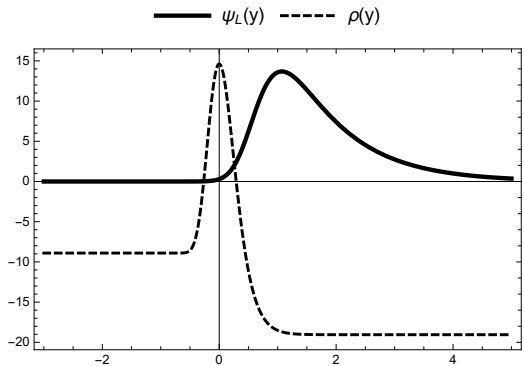}
\caption{Massless fermion $\psi_\text{L}$ shifted outside of energy density $\rho$, for $\lambda>\lambda_2$.}\label{Shift}
\end{figure}

Now, to simplify we will consider the case $\beta=3\alpha/e^2$, where the wall is centered in $y=0$, without loss of generality. However, the location of the fermion is shifted with respect to
the brane. To demonstrate this, let us calculate $\psi_{\text{L}}$ near $y=0$,
\begin{equation}
    \psi_{\text{L}}(y)\sim \exp\left[-\frac{\alpha}{2e\delta}(\lambda\phi_0-8\alpha)\left(y-y_0\right)^2\right]+{\cal O}(\alpha y/\delta)^3
\end{equation}
where
\begin{equation}\label{y0}
    y_0\simeq\frac{e\delta \lambda}{2\alpha}\left(\frac{\epsilon-\phi_0/e}{\lambda\phi_0-8\alpha}\right)\ll 1 .
\end{equation}
Thus, the wave function of the fermion is centered at $y_0$ with 
a width given by $\sqrt{e\delta}/\sqrt{ (\lambda\phi_0-8\alpha)\alpha}$. The fermion's shift was verified numerically beyond the approximation $\alpha y/\delta\ll 1$, and the results are shown in Fig.\ref{Shift}.

Unlike in this article, the shift  has previously been used to locate fermions on the wall: in \cite{Palma:2005xv}  where the spacetime is flat with an orbifold geometry $\mathbb{R}^4\times\text{S}^1$; and in \cite{Guerrero:2006gj} where the 5-dimensional spacetime is generated by a static double domain wall, in order to have the graviton and the fermion in different sub-walls.

For a scenario generated by a single wall with one non-compact extra dimension, shifting the fermion does not favor the creation of a brane world, since the maximum of probability is not on the wall. Therefore, in agreement with (\ref{y0}), $\epsilon=\phi_0/e$ is required to center the fermion on the wall and under this circumstance it is possible to evaluate the migratory effect.

From (\ref{lambdaY}) and (\ref{cota2}), the following is obtained 
\begin{equation}
   \lambda_{1-}=\frac{6\alpha}{e \phi_0}\quad , \qquad
   \lambda_{1+}=\frac{(e^2-3)}{(e-1)}\frac{2\alpha}{e \phi_0}  
\end{equation}
and 
\begin{equation}
    \lambda_2=\frac{8\alpha}{e \phi_0}
\end{equation}
in such a way that $\lambda_{2}>\lambda_{1-}>\lambda_{1+}$.

\begin{figure}[h!]
        \centering
        \begin{subfigure}{0.45\textwidth} 
            \includegraphics[width=\textwidth]{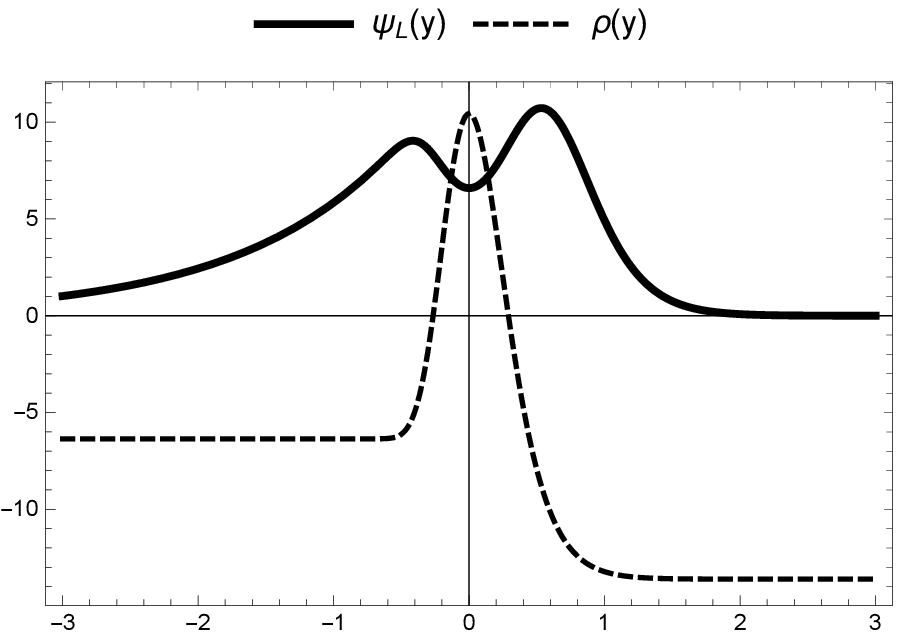}
        \end{subfigure}       
        \begin{subfigure}{0.45\textwidth} 
     \includegraphics[width=\textwidth]{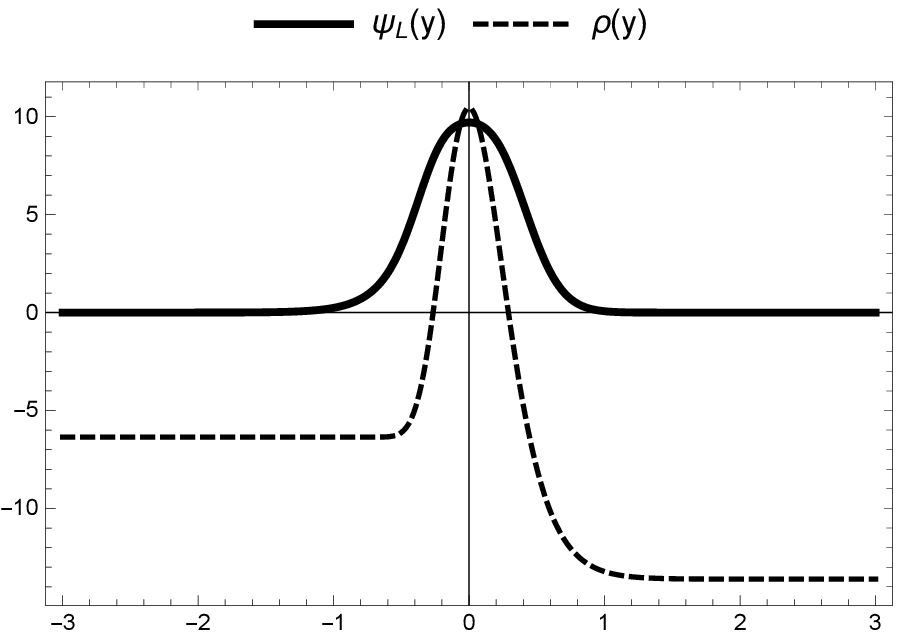}
        \end{subfigure}
        \caption{Fermion zero mode $\psi_{\text{L}}$ and energy density $\rho$  for $\lambda_1<\lambda<\lambda_2$ (left) and $\lambda>\lambda_2$ (right). }\label{plotrhoa}
    \end{figure}

In Fig. \ref{plotrhoa}, we plot the zero mode fermion for some values of $\lambda$. For the migratory sector, $\lambda_{1+}<\lambda<\lambda_2$, we can see two interfaces associated to the bounded state of $\psi_\text{L}$. The difference in the amplitude of the peaks is generated by the gravitational repulsion applied on the fermions, which expels them towards the region of greater curvature. For $\lambda>\lambda_2$ the probability's maximum is in $y=0$, and therefore, $\psi_\text{L}$ is confined on the wall. 

\section{Discussion}\label{S5}

The Yukawa coupling counteracts the repulsion exerted on the fermions by the gravitation of the wall, such that, utilizing a critical value of the Yukawa constant $\lambda>\lambda_1$ (obtained by asymptotic analysis of the wave function), ensures the normalization of  one of the chiral states. However, we found a range of values for the Yukawa constant, $\lambda_1<\lambda<\lambda_2$ (by the Laplace method), where the fermion is in a normalizable state orbiting around the wall.

We calculated the Yukawa bounds, $\lambda_1$ and $\lambda_2$, in two scenarios generated by a thick domain wall with or without reflection symmetry. 
  
On the domain wall solution with reflection symmetry \cite{Gremm:1999pj}, $A(y)$ and $\phi(y)$, the chiral massless fermion $\psi_\text{L}$ is normalizable around the wall. On the other hand,  the symmetric wall is part of a family of asymmetric solutions where $A(y)\rightarrow A(y)+\beta y$ and $\phi(y)\rightarrow \phi(y)-\epsilon$ \cite{Guerrero:2006gj}, and where the zero mode of the fermion is determined by
\begin{equation}\label{psiW}
    \psi_\text{L}(y)\rightarrow\psi_\text{L}(y)e^{-(2\beta-\lambda\epsilon)y} ,
\end{equation}
the fermion is shifted from the wall by the term $e^{-2\beta y}$. Therefore, it is not possible to evaluate the migratory effect on another member of the family of solutions, different than the $Z_2$-limit case: $(\beta, \epsilon)\rightarrow (0, 0)$. 

To evaluate the migratory effect on a scenario without $Z_2$ symmetry,  an intrinsically asymmetric domain wall \cite{CastilloFelisola:2004eg}, $A(y)$ and $\phi(y)$, is required. This scenario is also a member of a two-parameter family, $A(y)+\beta y$ and $ \phi(y)-\epsilon$, but without the $Z_2$-limit, and to ensure that $\psi_\text{L}(y)$ is centered in the maximum of $\rho(y)$, $\beta=3\alpha/e^2$ and $\epsilon=\phi_0/e$ must be utilized.

In both cases the constraints imposed on the fermion in order to get it on the wall, lead to a hierarchy  between $\lambda_2$ and $\lambda_1$, such that, regardless of the scenario's symmetry, the migratory effect could be present.

\section{Acknowledgements}
This work was supported by IDI-ESPOCH under the project titled {\it Domain walls and their gravitational effects.}

\end{document}